# spectrai: a deep learning framework for spectral data

Conor C. Horgan[1*] and Mads S. Bergholt[1*]

[1]Centre for Craniofacial and Regenerative Biology, King's College London, London SE1 9RT, UK

*Corresponding authors: mads.bergholt@kcl.ac.uk, conor.horgan@kcl.ac.uk

## Abstract

Deep learning computer vision techniques have achieved many successes in recent years across numerous imaging domains. However, the application of deep learning to spectral data remains a complex task due to the need for augmentation routines, specific architectures for spectral data, and significant memory requirements. Here we present spectrai, an open-source deep learning framework designed to facilitate the training of neural networks on spectral data and enable comparison between different methods. Spectrai provides numerous built-in spectral data pre-processing and augmentation methods, neural networks for spectral data including spectral (image) denoising, spectral (image) classification, spectral image segmentation, and spectral image super-resolution. Spectrai includes both command line and graphical user interfaces (GUI) designed to guide users through model and hyperparameter decisions for a wide range of applications.

## Introduction

Deep learning computer vision techniques have significantly advanced many fields of imaging, achieving state-of-the-art results across a variety of tasks including classification, segmentation, super-resolution, and denoising. For example, in medical imaging alone, deep learning has enabled incredible results for breast cancer prediction from mammography images[1], virtual histological staining of tissues[2], automated real-time colorectal polyp detection and segmentation[3], and deep learning-based super-resolution fluorescence microscopy[4].

The growing success of deep learning combined with the necessity for GPU computation and distributed training strategies to meet the demands of deep learning neural models led to the development of several deep learning libraries including cuDNN[5], Keras[6], Theano[7], Caffe[8], TensorFlow[9], and PyTorch[10]. Together, these libraries have helped to accelerate progress in deep learning, providing flexible and generalisable frameworks for building neural networks and the associated training and distribution pipelines. These libraries have made it easier to build and train neural networks across a wide variety of tasks from image classification to natural language processing and predictive modelling[11–14].

Due to their flexible nature, these deep learning libraries are somewhat domain agnostic, providing general purpose tools and functions that can be composed together to achieve domain-specific tasks. However, the successful application of deep learning to different applications requires significant expertise, with task-specific design decisions for network architecture, loss function, learning rate and schedule, as well as a multitude of hyperparameters. Achieving state-of-the art performance whilst preventing overfitting and undue model bias requires a careful understanding of the influence of each of these components, with tuning for a given task and dataset. For domains with non-standard data formats and processing requirements, significant domain-specific implementation is required[15,16]. This is particularly true for the application of deep learning to spectral data.

Spectral acquisition and imaging play an important role in machine vision, remote sensing, and biomedical imaging. Techniques such as multi or hyperspectral imaging, Raman spectroscopy, Fourier transform infra-red (FTIR) spectroscopy, and mass spectrometry imaging (MSI) have demonstrated utility across clinical diagnostics[17–19], environmental monitoring[20,21], and materials characterisation[22,23]. These techniques can produce information-rich 1D data (a spectrum, λ) or 3D data (a spectral hypercube, $x \times y \times \lambda$), with significant potential for a multitude of deep learning applications.

While there is much potential for spectral deep learning applications, existing deep learning frameworks and models for computer vision are largely oriented towards RGB images. Spectral data differs substantially from RGB images, however, and poses numerous requirements for which standard neural network architectures, data augmentations, and hyperparameter defaults are often unsuitable. Application of existing deep learning models to spectral datasets thus requires careful modification and adaptation to make training on spectral data possible and effective. For example, while spectral augmentations (e.g., spectral flipping or shifting) may be applied, standard image augmentations (e.g., brightness or contrast changes) may introduce unwanted spectral distortions. Similarly, although 2D convolutional neural networks (CNNs) may be extended to multi-channel hyperspectral images, many spectral deep learning applications employ 1D or 3D CNNs, necessitating modification of existing 2D CNN architectures or the development of novel task-specific architectures. Lastly, the large size of spectral image hypercubes poses significant memory constraints which may require modification of network architectures and training hyperparameters (e.g., batch size, patch size, scaling, data augmentation) to enable effective single- or multi-GPU training.

Despite these additional considerations posed by spectral data, deep learning has seen increasing application across multiple spectral imaging domains, achieving improved pixel-wise classification across hyperspectral imaging, mass spectrometry, and IR spectroscopy[24–28], spectral and hyperspectral image denoising/correction[29–32], and spectral image super-resolution[29,33,34] with results superior to classical machine learning algorithms typically applied to spectral data[35–37]. However, the limited support for domain-specific tasks (e.g., spectral imaging) in current deep learning libraries requires individual research groups to implement their own data processing, augmentation, and training pipelines. This results in a substantial duplication of efforts and limits effective comparison between different methods.

Recently, several efforts have aimed at developing domain-specific deep learning tools to simplify deep learning application in different fields, reduce duplications of effort, and enable robust comparisons. In medical imaging for example, data formats, sizes, and processing requirements differ significantly from those of RGB images, and a number of medical image-specific deep learning platforms have been developed including DLTK[38], NiftyNet[15], Eisen[39], TorchIO[16], MONAI[40], and pymia[41]. These platforms provide a host of functions suitable for the processing, augmentation, and training of medical images and have seen great successes in the medical imaging community. Similarly, frameworks such as Selene[42], pysster[43], and Kipoi[44] provide functionality for deep learning on biological sequence and genomic data. Similarly, there have been efforts to provide a repository of neural network implementations targeted towards pixel-wise classification of hyperspectral data[45]. However, to the best of our knowledge, no framework has been developed for general purpose application to spectral data.

Here, we present spectrai, an open-source, general purpose deep learning framework designed specifically for spectral data. Spectrai is built on the popular PyTorch library and includes baseline implementations of several networks for tasks including spectral (image) denoising, spectral (image) classification, spectral image segmentation, and spectral image super-resolution. In addition to a Python command line interface, spectrai provides a MATLAB graphical user interface (GUI) to guide users through deep learning training and inference procedures.

Spectrai provides 1) an easy-to-use framework, with Python and MATLAB interfaces to guide users; and 2) baseline implementations of deep learning spectral infrastructure including data pre-processing, data augmentation, and neural network architectures to reduce duplication of effort across research groups and enable effective comparisons between different methods.

**Results**

Spectrai addresses the lack of deep learning tools and frameworks designed for spectral data to lower the barriers to applying deep learning models to spectral data. Spectrai is designed for both non-expert users and experienced deep learning practitioners. The core of spectrai consists of a spectral deep learning framework developed in Python, using PyTorch, that can be run through a command line interface. Experienced deep learning practitioners can easily modify the existing code to implement additional data processing and augmentation methods and extend the library of spectral neural network architectures. In addition, an easy-to-use MATLAB GUI (Figure 1) interfaces the Python codebase, guiding users on model selection, suitable loss functions, suggested initial learning rates, and default hyperparameters based on the selected task (e.g., spectral image segmentation).

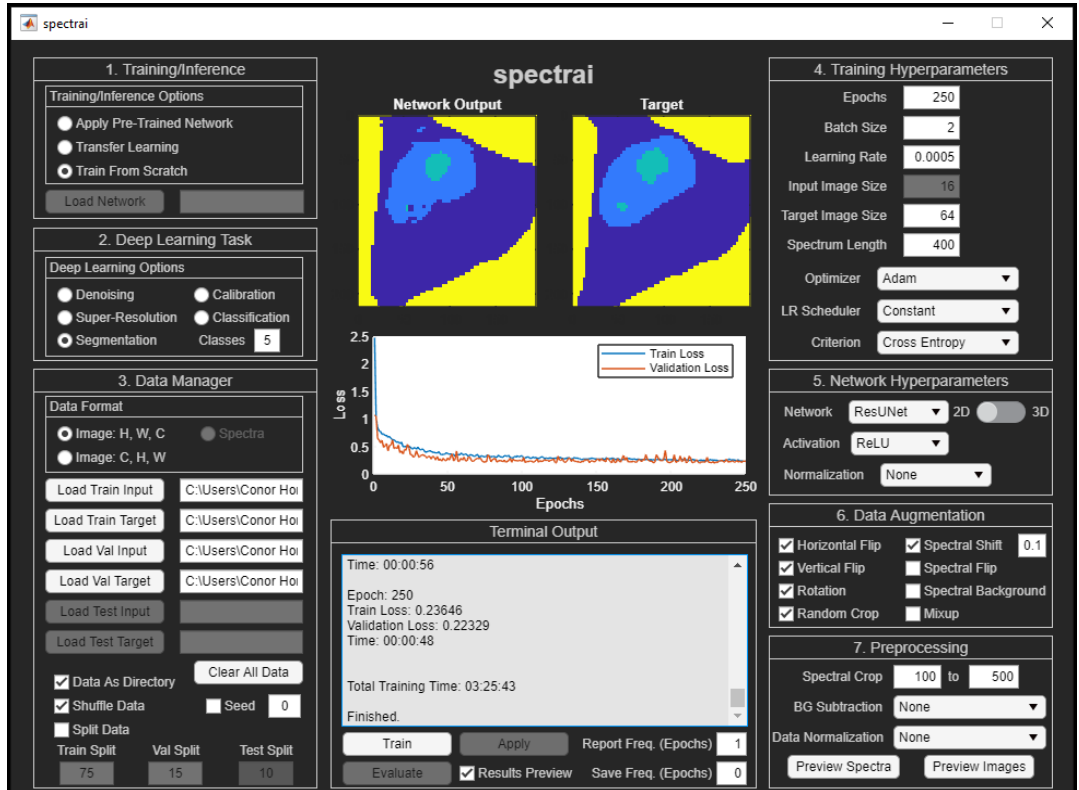

**Figure 1.** The spectrai MATLAB graphical user interface (GUI), applied here to segmentation of Raman spectroscopic images of MDA-MB-231 human breast cancer cells, simplifies neural network training for spectral data, guiding the user through task selection and data loading. The spectrai GUI applies task- and data-specific default hyperparameter values and restricts network and loss functions to those suitable for the selected task. Data loading, training hyperparameter, and data augmentation options not suitable for the selected task are similarly disabled to prevent unwanted errors.

Here we demonstrate three example applications of spectrai for deep learning of spectral data: spectral image segmentation, spectral denoising, and spectral image super-resolution.

*Spectral Image Segmentation*
Image segmentation is an important area of computer vision research with applications in fields as diverse as cell biology and remote sensing. Here, as with many areas of computer vision research, deep learning has demonstrated impressive results[46,47]. Together with the growing importance of spectral imaging across many fields, the development and application of neural networks for spectral image segmentation is essential.

To demonstrate the use of spectrai for spectral image segmentation, we used the recently published AeroRIT dataset[48]. This dataset consists of a single, large hyperspectral image (1973×3975 pixels) with reflectance data sampled every 10 nm from 400-900 nm (51 bands) (Note that the full dataset contains 372 bands between 387-1003 nm). To train a neural network for spectral image segmentation using spectrai we extracted 64×64-pixel non-overlapping patches, randomly splitting these into training, validation, and test sets (data split 85:10:5). Every pixel in the AeroRIT dataset images has been labelled as belonging to one of 6 classes (5 classes [roads, buildings, vegetation, cars, water] plus 1 class for unspecified pixels). Here we show segmentation results (Figure 2) achieved on the test set after training a UNet model with a batch size of 16 for 60 epochs, using the Adam optimizer[49] with a cross-entropy loss function and a constant learning rate of $1\times10^{-4}$ (training time: 22 minutes, Titan V GPU (NVIDIA)).

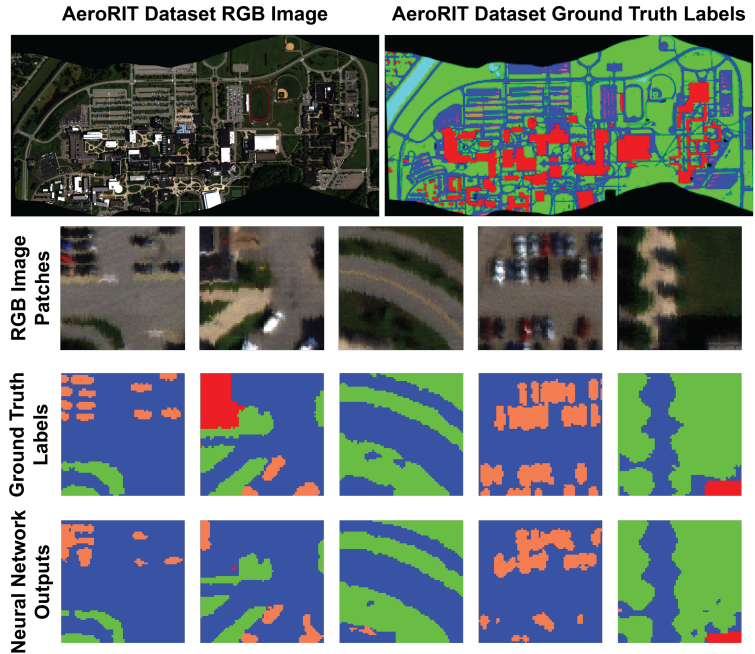

**Figure 2.** Spectrai semantic segmentation of AeroRIT hyperspectral remote sensing dataset[48] performed using a UNet architecture.

*Spectral Denoising*
Spectral denoising is an important task for the processing of spectral data, aiming to achieve the removal of unwanted noise artefacts while preserving important spectral information. Spectral denoising can for example be used to improve downstream data analysis and/or reduce data acquisition times.
Here, we illustrate the use of spectrai for the spectral denoising of a dataset of Raman spectra of MDA-MB-231 human breast cancer cells, recently developed as part of DeepeR towards efforts to improve Raman spectral acquisition times[29]. This dataset consists of 172,312 pairs of low SNR (0.1 s spectral integration time) and high SNR (1 s spectral integration time) spectra from 11 MDA-MB-231 cells. A ResUNet model was trained for 500 epochs using the Adam optimizer with an L1 loss function and a one-cycle learning rate scheduler (training time: 26 hours, Titan V GPU (NVIDIA)), achieving results superior to Savitzky-Golay filtering (Figure 3).

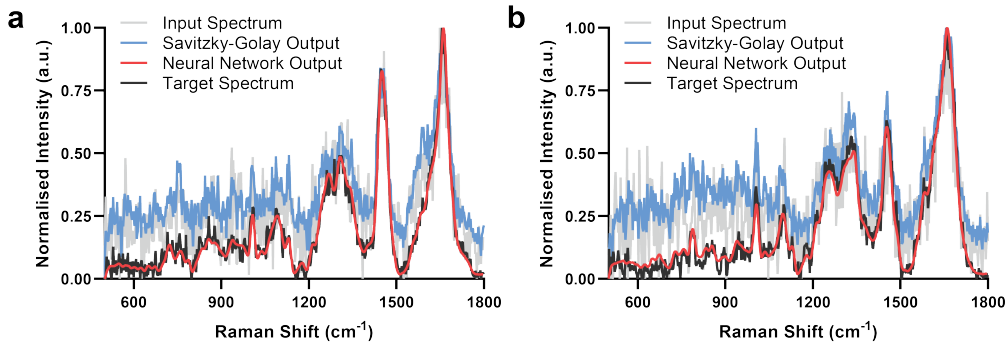

**Figure 3.** Spectrai spectral denoising of Raman spectral dataset of MDA-MB-231 breast cancer cells[29]. Example spectra shown for a) a lipid-rich cell region and b) for the cell nucleus.

*Spectral Image Super-Resolution*
Image super-resolution is yet another important task within computer vision research with a multitude of applications across a wide variety of domains. Spectral image-super resolution poses additional challenges relative to super-resolution of RGB images as both spatial and spectral information must be preserved.

For demonstration of spectral image super-resolution using spectrai, we focus on a dataset of intraoperative hyperspectral images of human brains for brain cancer detection[50]. The dataset consists of 36 hyperspectral images acquired intraoperatively during brain surgeries from 22 patients. The images are on average 439×400 pixels with 826 spectral bands between 400-1000 nm. A hyperspectral residual channel attention network (RCAN)[51] for 8x spatial super-resolution was trained on 64×64 pixel randomly cropped patches (bicubic downsampled 8x to produce 8x8 pixel inputs) from 33 of 36 images with spectral bands between 450-900 nm, with patches from one image reserved for the validation set and non-overlapping patches from a further image reserved for the test set (One hyperspectral image was removed from dataset as it was acquired from the same patient used for the test set). The RCAN model was trained with a batch size of 2 for 500 epochs using the Adam optimizer with an L1 loss function and a constant $1\times10^{-4}$ learning rate (training time: 17 hours, Titan V GPU (NVIDIA)), achieving results superior to bicubic upsampling (Figure 4).

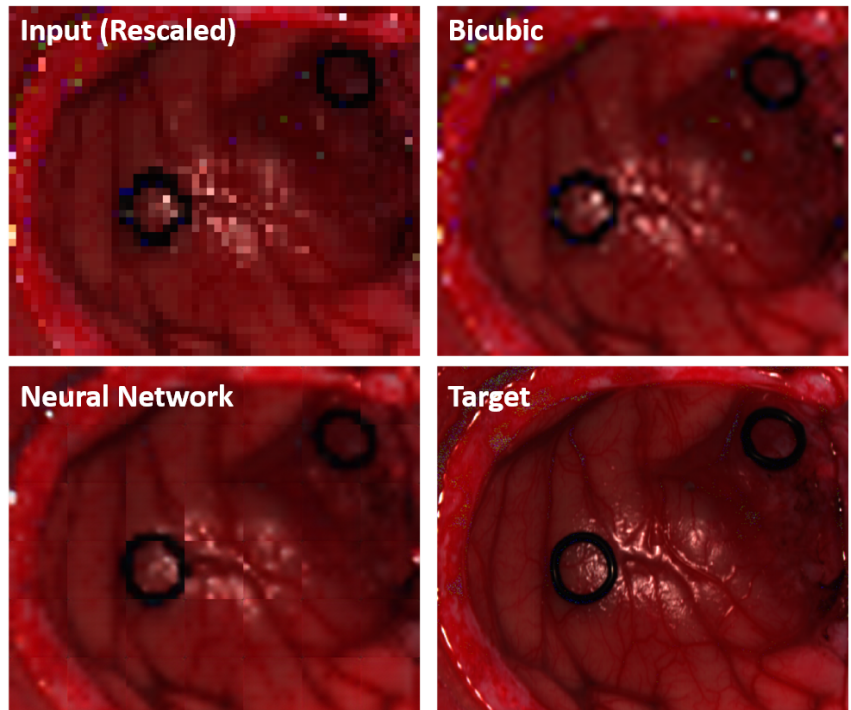

**Figure 4.** Spectrai 8x super-resolution of intraoperative hyperspectral brain image from the HELICoiD dataset[50] performed using a hyperspectral RCAN architecture.

## Discussion

Recent advances in the application of deep learning to spectral data have shown significant potential. However, the extension of deep learning methods to spectral data is non-trivial, requiring substantial overheads relative to the applications of deep learning to RGB images. Spectrai aims to minimise these overheads when applying deep learning to spectral data. By providing baseline implementations for spectral neural network architectures, spectral data augmentations, and the necessary infrastructure to train neural networks on spectral data, spectrai aims to make it quicker and easier for researchers to apply deep learning to new spectral datasets, compare models, and visualise results.

The core spectrai platform is built using Python and PyTorch, with open-source code designed to enable experienced practitioners to extend spectrai to introduce additional neural network models, data augmentations, and processing pipelines. The spectrai MATLAB GUI further provides an easy-to-use interface to the spectrai Python platform that guides users through task and model selection, data loading, dataset splitting, hyperparameter selection and data augmentations.

Spectrai code and documentation is available online and can be downloaded at https://github.com/conor-horgan/spectrai or installed as Python package using pip install spectrai. Spectrai is licensed under an open-source Apache 2.0 license.

Future development of spectrai aims to increase the range of applications, neural network architectures, and data augmentation and pre-processing methods available. Community feedback and contribution to the future development of spectrai is particularly welcomed. We hope to significantly expand the range of state-of-the-art spectral neural network architectures available to provide standard baselines for benchmarking and comparison purposes.

## Supporting Information

Spectrai code and documentation are available online at https://github.com/conor-horgan/spectrai.

## Acknowledgements
This work has received funding from the European Research Council (ERC) under the European Union's Horizon 2020 research and innovation programme (grant agreement No. 802778). This grant has allocated funding for open access publication.

## Conflict of Interest
The authors report no conflicts of interest.